\documentclass[twocolumn,twoside,slac_two]{revtex4}
\usepackage{graphicx}
\usepackage{fancyhdr}
\begin{document}

\title{Inclusive W/Z Production at CMS}

%

\author{P. Tan {\it on behalf of the CMS Collaboration}}
\affiliation{Fermi National Accelerator Laboratory, P.O. Box 500, 
Batavia, IL 60510, USA}

\begin{abstract}

At the LHC, the production cross sections of W/Z bosons are tens to 
hundreds of 
nanobarns. The production mechanism of these processes 
 is well established in the Standard Model and these processes 
can be used as ``standard 
candles'' to help commission the CMS 
detector for physics. Leptonic decays of W/Z bosons are expected to have 
very high trigger efficiency and signal to background ratio. Therefore they 
are ideal channels to study the properties of W/Z bosons in detail, 
such as cross sections and charge asymmetry. In this paper early CMS results 
on inclusive W/Z production at 10 TeV center-of-mass energy are discussed. 
\end{abstract}

\maketitle

\thispagestyle{fancy}


\section{Introduction}

The W and Z bosons were first discovered at CERN more than two decades ago~\cite{w, z}. Since
 then their properties have been extensively studied by different experiments 
to test the Standard Model~(SM) predictions and to 
explore the physics beyond-the-SM. At the Large Hadron Collider~(LHC)~\cite{lhc}, W and Z bosons will 
be produced with large rates. A large set of W/Z bosons will help us with 
detector commissioning initially and enable us to perform a 
large variety of W/Z physics studies with early LHC data.

The production mechanism of 
W/Z bosons at the LHC is well known. Higher order predictions of 
many W/Z observables have been carried out. For example, 
a recent next-to-leading order  
calculation~\cite{campbell} of total W and Z cross sections predicted 
that the cross sections are tens to hundreds of nanobarns 
for Z and W bosons, respectively.  
A calculation of differential cross section as a function of boson rapidity
 at next-to-next-leading 
order has also been carried out by C.~Anastasiou 
{\it et al.}~\cite{anastasiou}. 
In these theoretical predictions, 
errors due to the Parton Distribution Functions~(PDF) dominated total 
theoretical errors. The PDF error could be partially canceled out if we 
study ratios of cross sections, such as the lepton charge asymmetry 
 between 
$W^+$ and $W^-$ production, which is defined to be,
 \begin{equation}
A(\eta) = \frac{ \frac{d\sigma}{d\eta} (W^+\rightarrow l\nu) -  \frac{d\sigma}{d\eta} (W^-\rightarrow l \nu) }{  \frac{d\sigma}{d\eta} (W^+\rightarrow l\nu) +  \frac{d\sigma}{d\eta} (W^-\rightarrow l\nu) }.
\label{eq:asym}
\end{equation}
This charge asymmetry
 probes the valence-sea quark ratio in protons. Measurements 
of these observables at the LHC will enable us to test higher 
order calculations and provide new insights into 
proton structure.

\section{CMS Detector}

The Compact Muon Solenoid~(CMS) experiment is a 4$\pi$ general-purpose 
hadron-collider detector, which is suitable for high-$p_T$
physics studies 
at the LHC. 
The central feature of the Compact Muon Solenoid apparatus is a 
superconducting solenoid, of 6~m internal diameter, providing a field of 
3.8~T. Within the field volume are the silicon pixel and strip tracker, the 
crystal electromagnetic calorimeter (ECAL) and the brass/scintillator hadronic
 calorimeter (HCAL). Muons are measured in gaseous detectors embedded in the 
iron return yoke. Besides the barrel and endcap detectors, CMS has extensive 
forward calorimetry. 
The ECAL has an energy resolution of better than 0.5\,\% above 100~GeV. The HCAL, when combined with the ECAL, measures jets with a resolution $\Delta E/E \approx 100\,\% / \sqrt{E} \oplus 5\,\%$. The calorimeter cells are grouped in projective towers, of granularity $\Delta \eta \times \Delta \phi = 0.087\times0.087$ at central rapidities and $0.175\times0.175$ at forward rapidities. 
The muons are measured in the pseudorapidity window $|\eta|< 2.4$, with detection planes made of three technologies: Drift Tubes, Cathode Strip Chambers, and Resistive Plate Chambers. Matching the muons to the tracks measured in the silicon tracker results in a transverse momentum resolution between 1 and 5\,\%, for $p_{\rm T}$ values up to 1~TeV/$c$. 
The first level of the CMS trigger system, composed of custom hardware
 processors, uses information from the calorimeters and muon detectors
 to select (in less than 1~$\mu$s) the most interesting events (only one
 bunch crossing in 1000). The High Level Trigger 
processor farm further decreases the event rate from 100~kHz to 100~Hz, 
before data storage. 
A much more detailed description of CMS can be found elsewhere~\cite{cms}.

\section{Event Simulation}

The Monte Carlo~(MC) simulation used in the following studies was generated 
with the Pythia~\cite{pythia} event generator, where the CTEQ5L~\cite{cteq5l} 
PDF model was 
used. The center-of-mass energy 
was assumed to be 10 TeV. The generated events were then passed through 
the full CMS detector simulation with GEANT4~\cite{geant4}. Physics 
objects such as muons and electrons 
were reconstructed with standard CMS offline reconstruction sequence. 
The missing transverse energy~(MET) was reconstructed using energy deposits 
in CMS calorimeters. 

\section{Inclusive W Boson Cross Section}

At the LHC, leptonic decays of W/Z bosons were used to study the properties of 
W/Z bosons. The experimental signature of a W boson is a high-$p_T$ 
lepton and 
large MET due to presence of a neutrino in the final state. 
CMS conducted analyses to measure the inclusive W boson 
cross section in both muon and electron decays~\cite{muon, electron}. 

\begin{figure*}[bhtp]
\centering
\includegraphics[width=80mm, height=65mm]{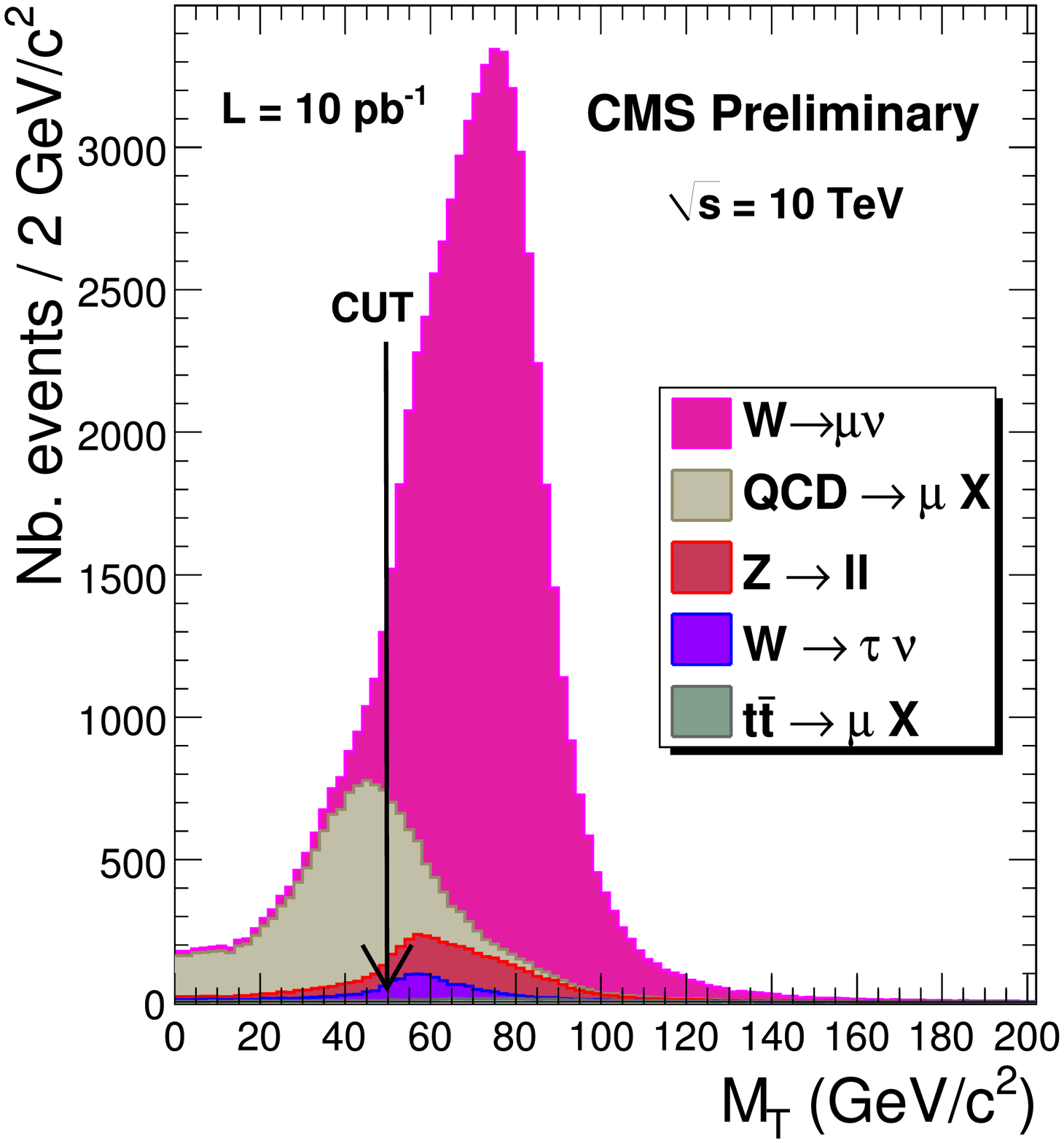}
\hfill
\includegraphics[width=80mm, height=65mm]{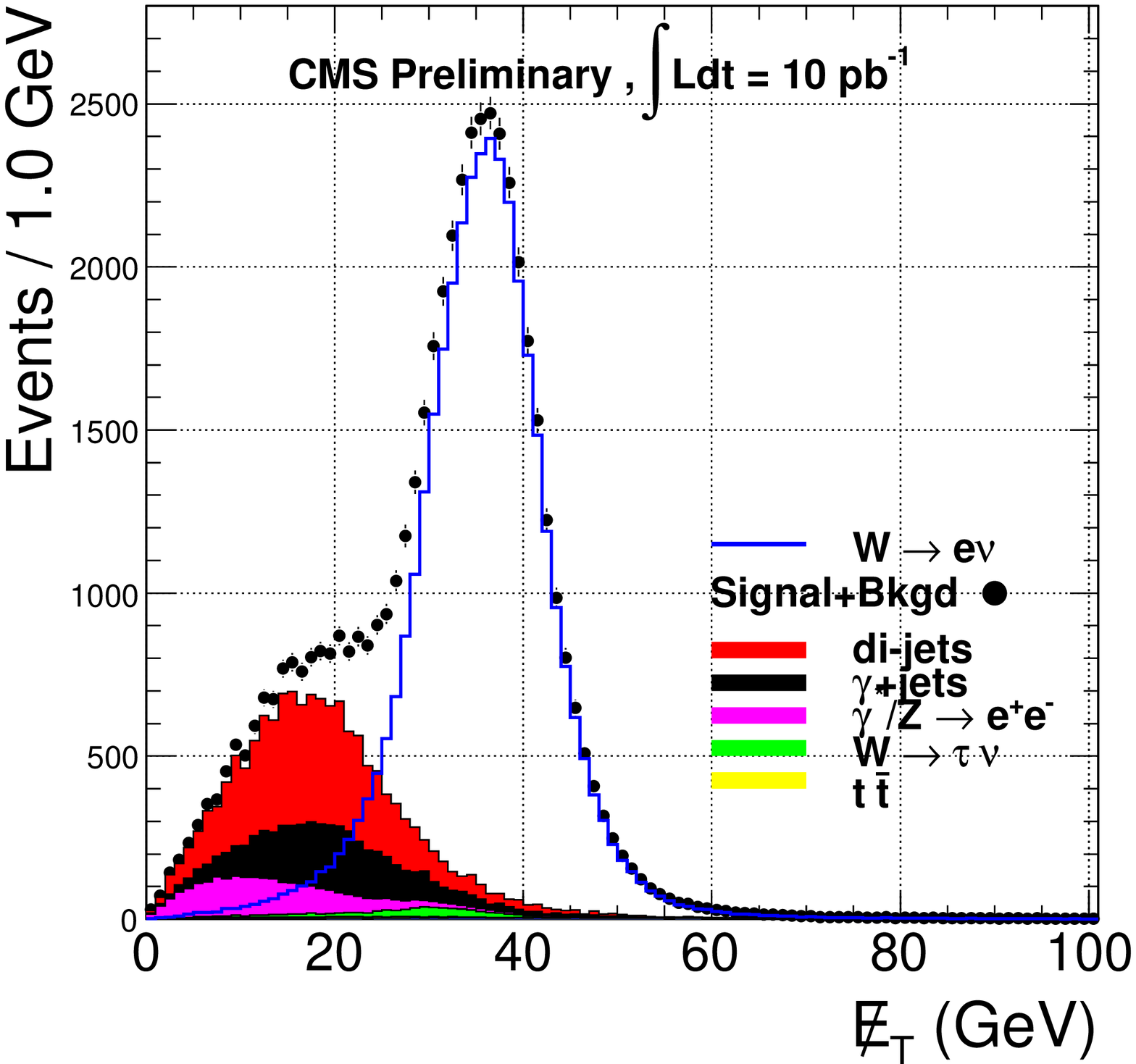}
\caption{ Left) The reconstructed $W\rightarrow \mu\nu$ transverse mass distribution. Right) The reconstructed $W \rightarrow e \nu$ MET 
 distribution. Both are 
normalized to 10 pb$^{-1}$ of integrated luminosity. The arrow in the 
left figure indicates that a cut on $m_T>$ 50 GeV was applied to select 
final data sample. } 
\label{fig:wxsec}
\end{figure*}

The trigger used in $W\rightarrow \mu\nu$ analysis is a single muon 
trigger with a minimum $p_T$ threshold of 15 GeV. The efficiency is 
above 90\%. The CMS single muon trigger system has coverage up to 
a pseudorapidity of $|\eta|<$ 2.1. 
The selection of $W\rightarrow \mu\nu$
candidates was done by first requiring an
 isolated muon with $p_T>$ 25 GeV and 
muon pseudorapidity $|\eta| <$2.0. Here the isolation is the $p_T$ sum 
of all tracks in a cone of radius 0.3 around the muon 
direction, normalized to  the muon 
$p_T$. This normalized isolation is required to be less than 0.09. The QCD 
dijet background was largely suppressed by the isolation requirement. 
Other processes, such as Drell-Yan, $t\bar{t}$, $W\rightarrow \tau\nu$, 
could 
also fake a $W\rightarrow \mu\nu$ event. Background events were further 
suppressed by requiring transverse 
mass~\footnote{Here $m_T=\sqrt{2 \cdot p_T \cdot MET \cdot (1-cos(\Delta \phi_{\mu, MET}))}$, where $\Delta \phi_{\mu, MET}$ is the angular difference 
between muon and missing transverse moment in the plane transverse to the 
beam direction. } $m_T>50$ GeV, as shown in Fig.~\ref{fig:wxsec}, where 
the expected $W\rightarrow\mu\nu$ signal and background events for 
an integrated luminosity of 10 pb$^{-1}$ is shown. 

CMS also studied the inclusive W boson cross section in electron decays. 
A single electron trigger was  
used, which has an efficiency of about 97\%. An electron candidate was  
required to have transverse energy~($E_T$) deposit in the CMS ECAL 
detector $E_T>$ 30 GeV and $|\eta|<$2.5. The shower shape of a electron 
candidate was also required to be consistent with an 
 electromagnetic interaction. 
Comparing to the muon channel analysis, in addition to 
background processes such as, QCD dijet production, Drell-Yan, $t\bar{t}$,
$W\rightarrow \tau\nu$, photon plus jet production also contributes. 
The QCD dijet background was significantly reduced by an isolation 
requirement. Here isolation was computed using transverse components 
of energy deposits in CMS calorimeters and tracks in a cone of radius 0.4 
around the electron direction. Figure~\ref{fig:wxsec} shows the 
reconstructed MET distribution for $W\rightarrow e\nu$ signal and 
background events after all event selections were applied.

The $W\rightarrow l \nu$ cross section is related 
to the background subtracted  number of signal 
events, $N_W$,
\begin{equation}
\sigma_W\times BR(W\rightarrow l\nu) = \frac{N_W}{A_{W}\times \epsilon_W \times {\cal L}},
\label{eq:xsec}
\end{equation}
where $A_W$ is acceptance for W signal events, $\epsilon_W$ is the 
W reconstruction and selection efficiency, and ${\cal L}$ is the integrated 
luminosity. While acceptance has to be estimated with MC, the 
lepton reconstruction and selection efficiency can be derived directly 
from data with a tag-and-probe method~\cite{tag_probe}. The expected 
statistical error 
of the measured cross section is 1.5\% for an integrated luminosity 
of 10 pb$^{-1}$. The systematic error is expected to be dominated by the 
luminosity error, which is expected to be about 10\% at the CMS 
start-up~\cite{lumi}.

\section{Inclusive Z Boson Cross Section}

\begin{figure*}[hbtp]
\centering
\includegraphics[width=80mm, height=65mm]{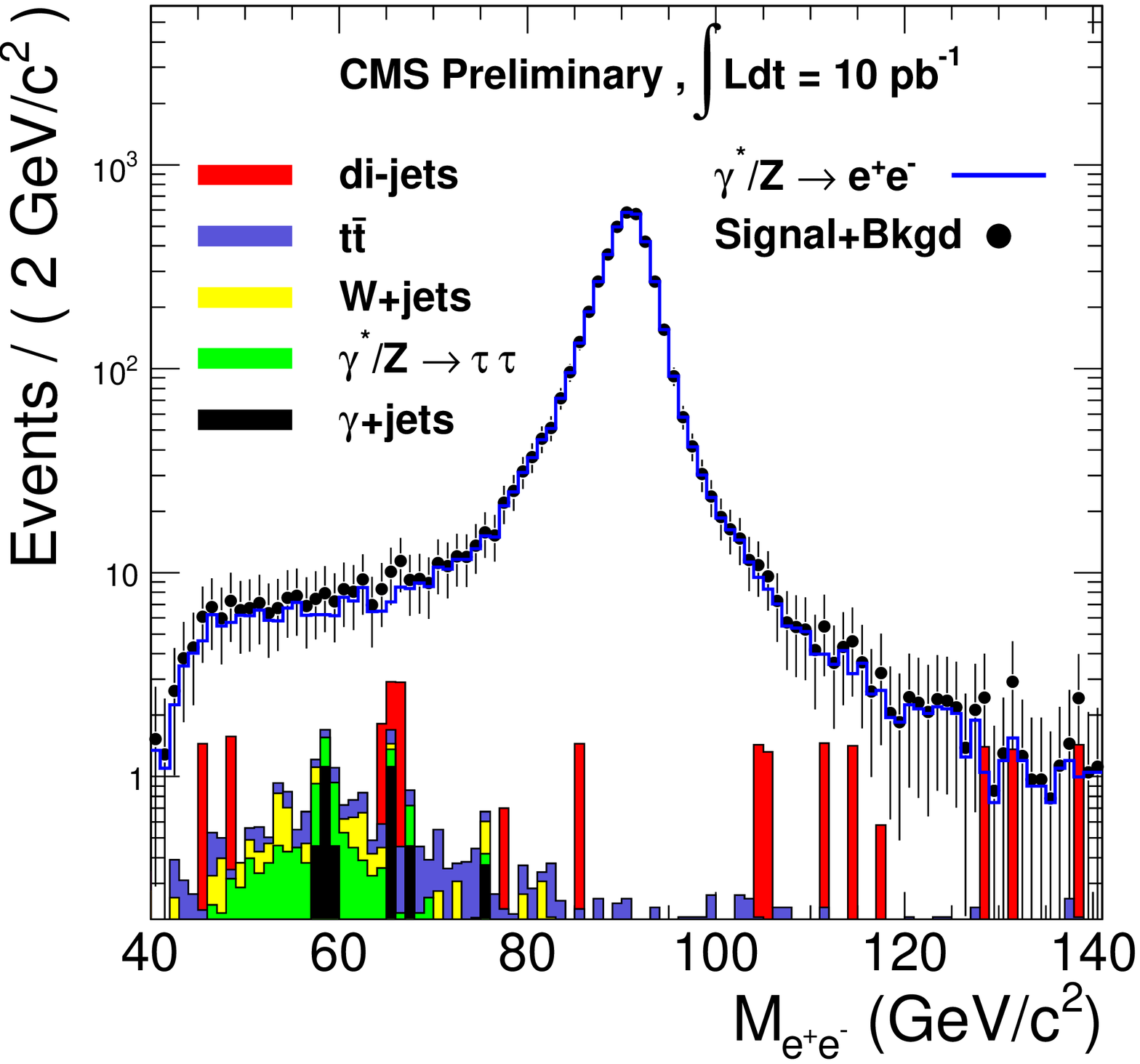}
\hfill
\includegraphics[width=80mm, height=65mm]{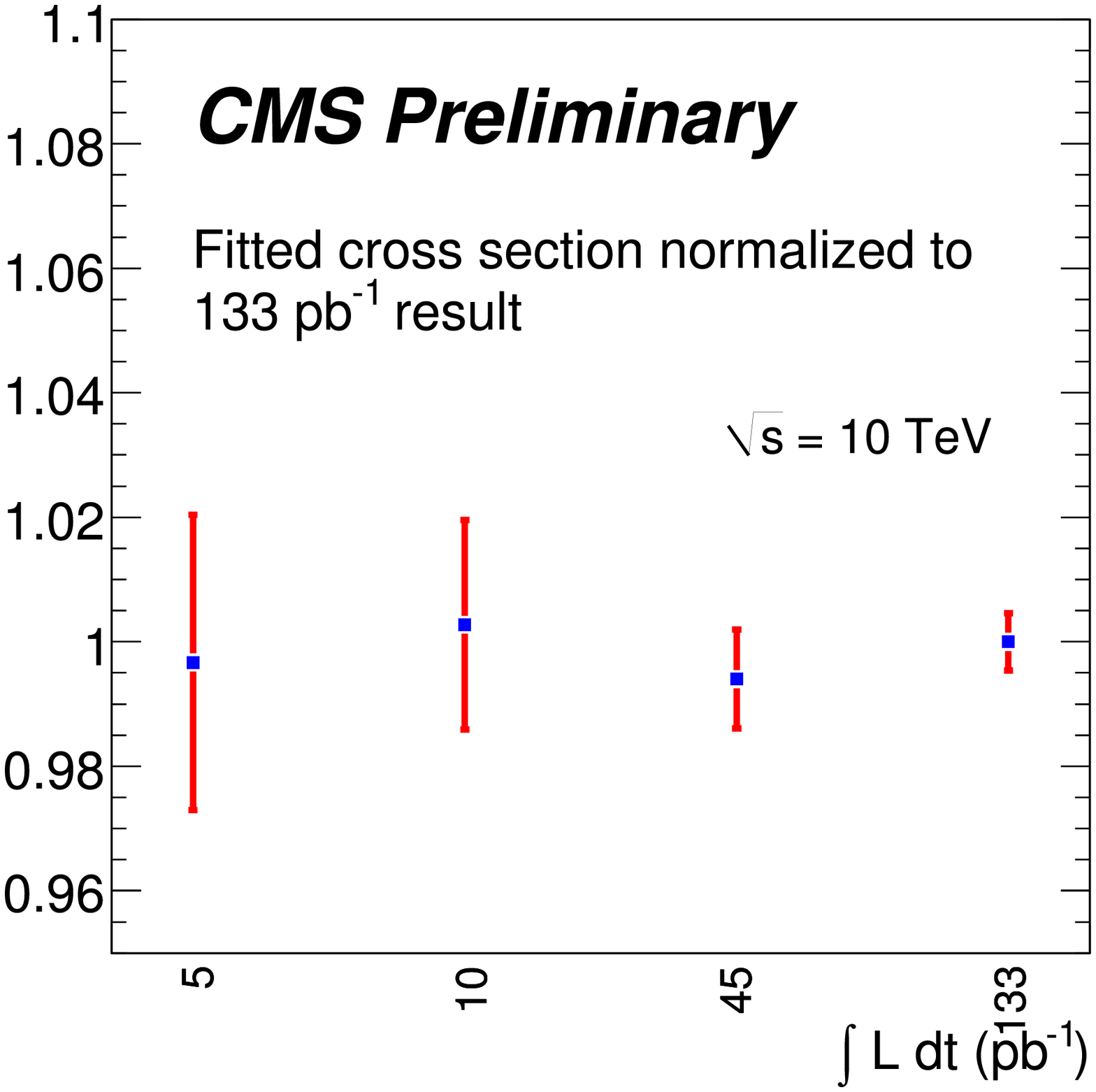}
\caption{Left) The reconstructed $Z\rightarrow ee$ invariant mass distribution for an integrated luminosity of 10 pb$^{-1}$. Right) The $Z \rightarrow \mu\mu $ cross 
section as a function of MC statistics corresponding to different luminosity scenarios. The results were normalized to the cross section determined with  
 133 pb$^{-1}$ of integrated luminosity.  Only statistical errors are shown.} 
\label{fig:zxsec}
\end{figure*}

Similarly the inclusive Z 
boson production is another ``standard candle'' to help us commision CMS 
for physics. 
CMS studied experimental sensitivities to the  inclusive Z boson cross section 
in both di-muon and di-electron decays~\cite{muon, electron}. 
Comparing to the $W\rightarrow l\nu$ analysis, due to presence of 
two isolated high-$p_T$ leptons the background was expected to be less 
than one percent after final event selection. The major background 
contributions were from QCD dijet, W plus jets, $t\bar{t}$, $Z\rightarrow \tau\tau$. Figure~\ref{fig:zxsec} shows the  reconstructed $Z\rightarrow ee$ invariant mass distribution for an integrated luminosity of 10 pb$^{-1}$. 
Both expected signal and background contributions are shown. The final $Z\rightarrow e e $ sample was selected from events with 70~GeV~$<m_{ee}<$110~GeV. 
The cross sections were obtained after correcting for efficiencies and 
acceptance following Eq.~\ref{eq:xsec}. The expected 
cross section for $Z \rightarrow \mu\mu $ decays as a function of 
statistics corresponding to different luminosity scenarios is also 
shown in Fig.~\ref{fig:zxsec}. The results were normalized to the cross 
section determined 
with a MC sample corresponding to an integrated luminosity of  133 pb$^{-1}$.
 The expected statistical error at 10 pb$^{-1}$ of integrated luminosity is 
about 2\%. Both analyses showed similar sensitivities. The 10\% luminosity 
uncertainty is again expected to dominate the total error.  

\section{Constraints to Parton Distribution Functions}

\begin{figure*}[t]
\centering
\includegraphics[width=80mm, height=65mm]{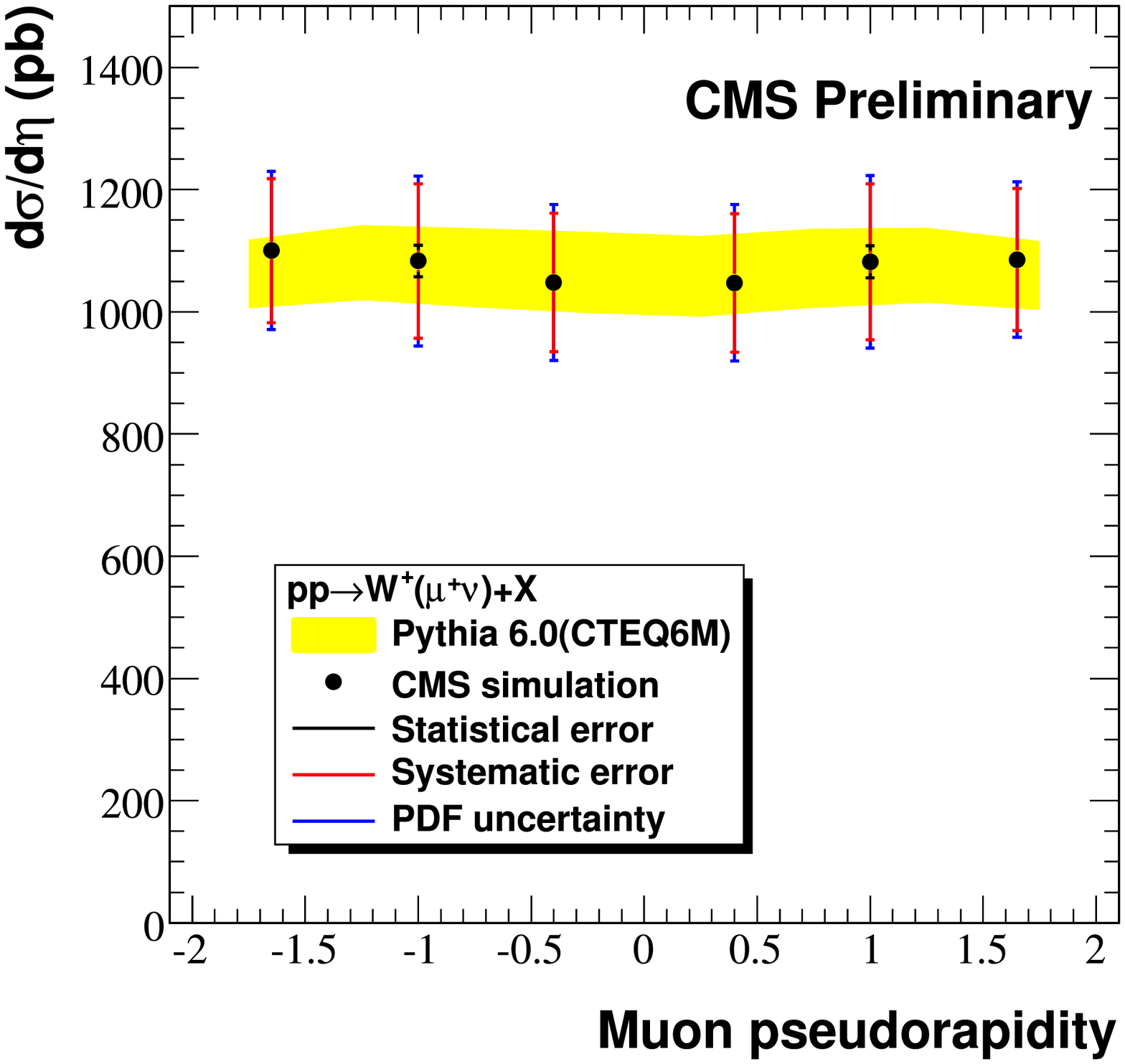}
\hfill
\includegraphics[width=80mm, height=65mm]{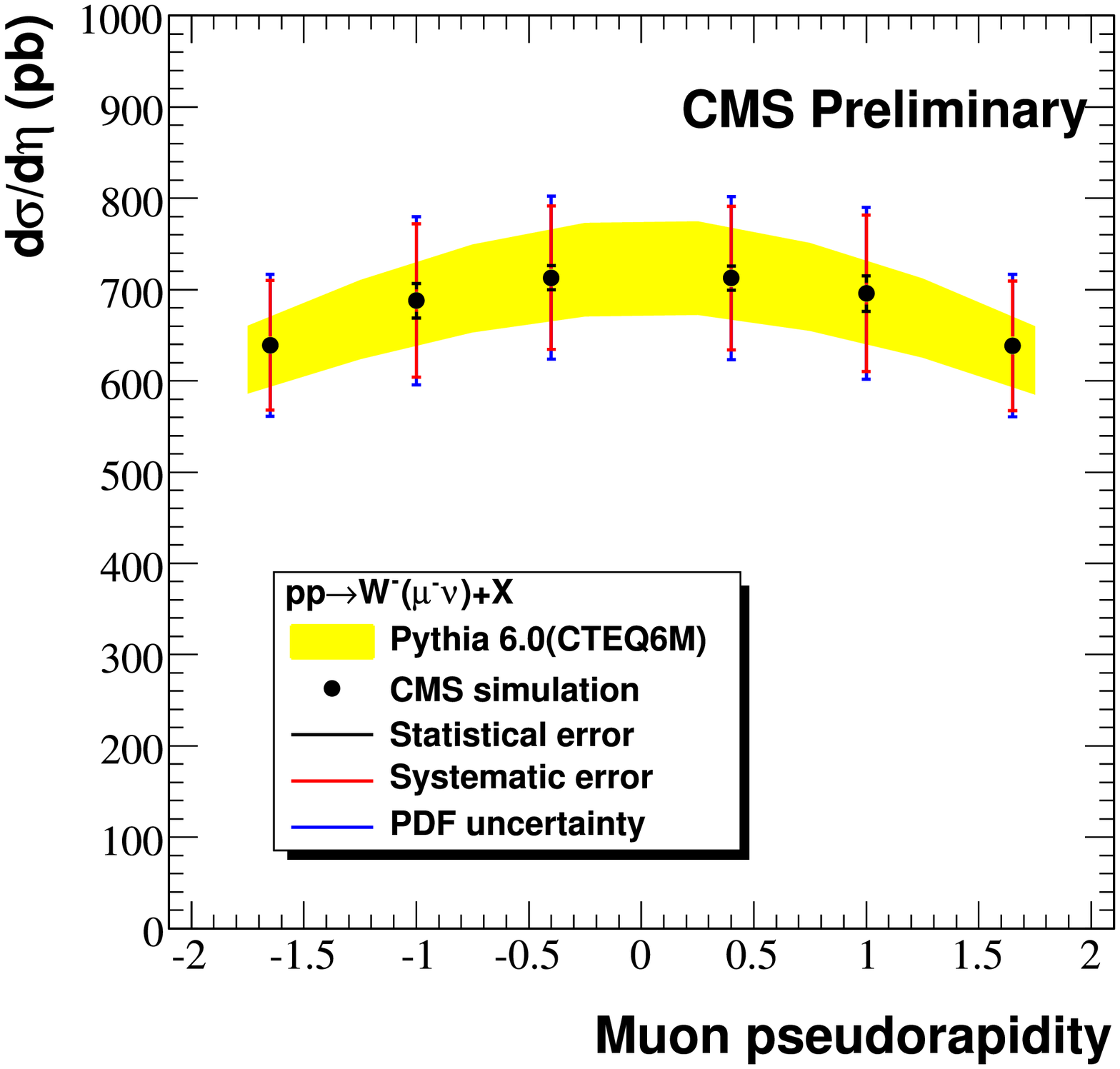}
\caption{The expected $W \rightarrow \mu\nu$ differential cross sections for 
an integrated luminosity of 10 pb$^{-1}$. Left) $\mu^+$, and Right) $\mu^-$. } 
\label{fig:muondiffxsec}
\end{figure*}

The large W/Z cross sections at the LHC makes high-precision 
differential measurements possible. CMS performed a measurement of the 
muon differential cross section measurement as a function of muon 
pseudorapidity in  
inclusive $W\rightarrow \mu\nu$ production~\cite{asym}. 
This analysis utilized the same trigger 
path as the inclusive $W\rightarrow\mu\nu$ cross section 
analysis. The $W\rightarrow \mu\nu$
candidates were selected with muon $p_T>25$ GeV and 
MET$>$20 GeV. A calorimeter-based isolation was used to suppress 
the QCD dijet contribution. After correcting for efficiencies and 
acceptance, the expected muon pseudorapidity distributions 
at an integrated luminosity of 10 pb$^{-1}$
 are shown in Fig.~\ref{fig:muondiffxsec}. 
The PDF error on the experimental data points is the theoretical error 
in estimating acceptance. Among other systematic errors, the 10\%
 luminosity error dominates. 
The results were compared to theoretical 
predictions from Pythia. We estimated the 
PDF error using the CTEQ6M~\cite{cteq6m} PDF model with 
PDF-reweighting technique~\cite{bourilkov}. With 10 pb$^{-1}$ of integrated luminosity, the results are still dominated by experimental 
systematic error. 

\begin{figure}[h]
\centering
\includegraphics[width=80mm]{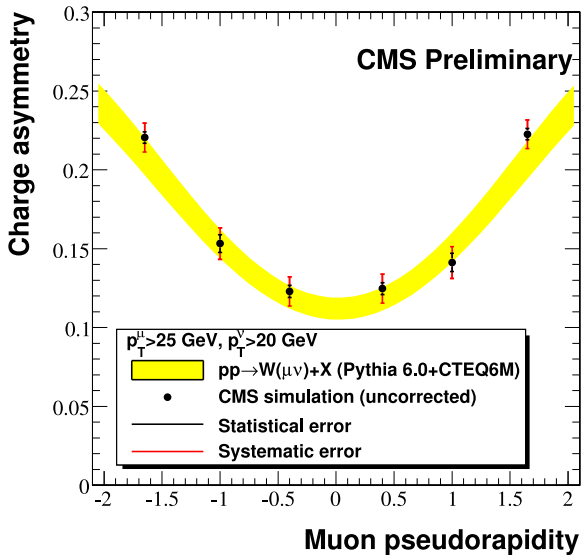}
\caption{The expected muon charge asymmetry for 100 pb$^{-1}$ of simulated 
luminosity.  
The systematic error is dominated by the statistical error on the efficiency
 ratio between $\mu^+$ and $\mu^-$ determined using 100 pb$^{-1}$ of
 Drell-Yan MC. } 
\label{fig:wasym}
\end{figure}

To minimize experimental systematic errors, CMS also studied the 
muon charge asymmetry defined in Eq.~\ref{eq:asym} because 
many experimental errors cancel out. The PDF error on this quantity 
is only at few percent level~\cite{mandy}. 
The analysis strategy 
was identical to the muon differential cross section analysis described above.
 We computed 
the observed charge asymmetry, $A^{obs.}(\eta)$, with background subtracted 
 number of $W\rightarrow \mu\nu$ 
signal events, $N^{W\rightarrow \mu\nu} (\eta)$,
\begin{equation}
A^{obs.}(\eta) = \frac{ N^{W^{+}\rightarrow \mu^+ \nu }(\eta) -N^{W^{-}\rightarrow \mu^- \nu }(\eta) }{ N^{W^{+}\rightarrow \mu^+ \nu }(\eta) + N^{W^{-}\rightarrow \mu^- \nu }(\eta)  },
\label{eq:asym:obs}
\end{equation}
assuming that reconstruction and selection efficiency ratios between $\mu^+$
 and $\mu^-$ are one. Because of the weak decay of W bosons,  the 
acceptance ratio between $\mu^+$
 and $\mu^-$ differs from unity. In this analysis, we did not correct for 
the acceptance difference in $A^{obs.}(\eta)$ but absorbed  it into the 
 theoretical 
predictions. The charge asymmetry as a function of muon 
pseudorapidity for an 
integrated luminosity of 100 pb$^{-1}$ is shown in 
Fig.~\ref{fig:wasym}~\cite{asym}. The result is compared to theoretical 
predictions from Pythia, where the PDF error was estimated with CTEQ6M PDF 
model. 
The systematic error is dominated by the 
statistical error on the 
efficiency ratio between $\mu^+$ and $\mu^-$ determined using 100 pb$^{-1}$ 
of Drell-Yan MC. Even with a conservative treatment of systematic 
errors, the total errors of this measurement are comparable to the PDF errors 
and potentially could provide new constraints on different PDF models.  

The Z boson differential cross section as a function of Z rapidity
can also be used to test higher order perturbation calculations and 
put constraints on PDF models. CMS performed a study of this 
observable using $Z\rightarrow ee $ events~\cite{zeeshape}.  To remove the 
luminosity uncertainty in this measurement, the following observable 
was studied,
\begin{equation}
\frac{1}{\sigma} \cdot \times \frac{ d\sigma(Z\rightarrow ee) }{ dy_i } = 
\frac{ \sum_i{ (\epsilon \times A)_i} }{ \sum_i{N_i} } \cdot \frac{ N_i }{  \Delta_i (\epsilon \times A)_i  },
\label{eq:zshape}
\end{equation}
where for each bin $i$ of rapidity ($y_i$), $N_i$ is the number of background subtracted $Z\rightarrow ee $  candidates, $\Delta_i$ is the bin width, 
and $(\epsilon \times A)_i$ is the product of the efficiency and acceptance for detecting and reconstructing a Z boson with rapidity $y_i$.  
 
With conventional electron reconstruction at the CMS, where 
both CMS ECAL and tracking system are utilized, the coverage for electrons is  
up to pseudorapidity of about 2.5. The reconstructed $Z\rightarrow ee $  
candidates can be used to directly probe partons with kinematics 
~\footnote{Here the initial state parton momentum fraction 
$x_{1,2}=\frac{ M_Z }{ \sqrt{s} } \cdot e^{\pm y}$,
where $s$ is the center of mass energy at the LHC. }
outside the range of previous experiments~\cite{campbell}. The kinematics
reach was further extended by using electrons reconstructed 
using the CMS Forward Hadronic~(HF) calorimeter~\cite{hf}, which has a  
coverage up to pseudorapidity of 4.6. The shower shape of HF reconstructed 
electron candidates was utilized to remove  $Z\rightarrow ee $ 
background events effectively. 

\begin{figure}[h]
\centering
\includegraphics[width=80mm]{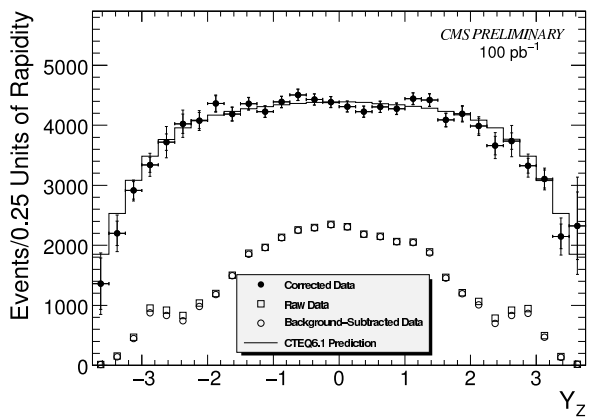}
\caption{The expected Z boson rapidity shape for an integrated luminosity of 100 pb$^{-1}$.} 
\label{fig:zeeshape}
\end{figure}

The final results for the rapidity measurement for an integrated 
luminosity of 100 pb$^{-1}$ is shown in Fig.~\ref{fig:zeeshape}. The 
background in the HF region is well under 
control.  
The expected measurements are compared to predictions with CTEQ6.1 PDF model~\cite{cteq61}.  With an integrated luminosity of 100 pb$^{-1}$, we are expecting 
to provide new constraints on different PDF models. 

\section{Summary}

CMS performed MC studies to explore
W/Z boson production with initial LHC data.  With 10 pb$^{-1}$ of 
integrated luminosity, the inclusive W/Z cross sections could be 
established with 1-2\% statistical precision. However, it is expected 
that the luminosity uncertainty will dominate the total error in 
these measurements. 

The sensitivities to constrain different PDF models at CMS were also 
explored using 
measurements of the muon charge asymmetry in inclusive $W \rightarrow \mu\nu$
process and the Z rapidity shape in inclusive $Z\rightarrow ee$ 
process. With about 100 pb$^{-1}$ of integrated luminosity, both 
measurements could provide 
constraints on different PDF models.

\begin{acknowledgments}

We thank the technical and administrative staff at CERN and other CMS 
Institutes, and acknowledge support from: FMSR~(Austria); FNRS and 
FWO~(Belgium); CNPq, CAPES, FAPERJ, and FAPESP~(Brazil); MES~(Bulgaria); 
CERN; 
CAS, MoST, and NSFC~(China); COLCIENCIAS~(Colombia); MSES~(Croatia); 
RPF~(Cyprus);
 Academy of Sciences and NICPB~(Estonia); Academy of Finland, ME, and 
HIP~(Finland); CEA and CNRS/IN2P3~(France); BMBF, DFG, and HGF~(Germany); 
GSRT~(Greece); OTKA and NKTH~(Hungary); DAE and DST~(India); IPM~(Iran); 
SFI~(Ireland); INFN~(Italy); NRF~(Korea); LAS~(Lithuania); CINVESTAV, 
CONACYT, SEP, and UASLP-FAI~(Mexico); PAEC~(Pakistan); SCSR~(Poland); 
FCT~(Portugal); JINR~(Armenia, Belarus, Georgia, Ukraine, Uzbekistan); 
MST and MAE~(Russia); MSTDS~(Serbia); MICINN and CPAN~(Spain); Swiss Funding
 Agencies~(Switzerland); NSC~(Taipei); TUBITAK and TAEK~(Turkey);
 STFC~(United Kingdom); DOE and NSF~(USA). Individuals have received 
support from the Marie-Curie IEF program~(European Union); the Leventis 
Foundation; the A. P. Sloan Foundation; the Alexander von Humboldt Foundation. 

\end{acknowledgments}

\bigskip 

\end{document}